\documentclass[aps,prl,twocolumn,showpacs]{revtex4}

\usepackage{graphicx}
\usepackage{latexsym} 
\usepackage{amsmath,amscd}
\usepackage{times}

\begin{document}

\title{\bf\noindent Large Deviations of Extreme Eigenvalues of Random Matrices}
\author{David S. Dean$^1$ and Satya N. Majumdar$^2$}
\affiliation{
$^1$ Laboratoire de Physique Th\'eorique (UMR 5152 du CNRS),
Universit\'e Paul Sabatier, 118, route de Narbonne, 31062
Toulouse Cedex 4, France\\
$^2$ Laboratoire de Physique Th\'eorique et 
Mod\`eles Statistiques (UMR 8626 du CNRS), 
Universit\'e Paris-Sud, B\^at. 100, 91405 Orsay Cedex, France}

\begin{abstract}

We calculate analytically the probability of large deviations
from its mean of the largest (smallest) eigenvalue of
random matrices belonging to the Gaussian orthogonal, unitary
and symplectic ensembles. In particular, we show
that the probability that all the eigenvalues of an $(N\times N)$
random matrix are positive (negative) decreases for large $N$ 
as $\sim \exp[-\beta \theta(0) N^2]$
where the parameter $\beta$ characterizes the ensemble and the exponent
$\theta(0)=(\ln 3)/4=0.274653\dots$ is universal. We also calculate exactly
the average density of states in matrices whose eigenvalues are restricted to
be larger than a fixed number $\zeta$, thus generalizing the celebrated 
Wigner semi-circle law. The density of states generically
exhibits an inverse square-root singularity at $\zeta$.

\end{abstract}
\maketitle

\begin{figure}
\includegraphics[width=.9\hsize]{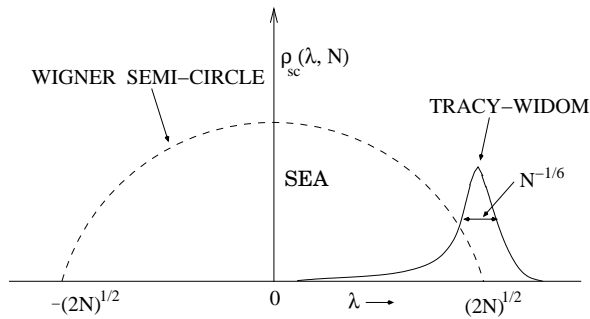}
\caption{The dashed line shows the semi-circular form of the
average density of states. The largest eigenvalue is centered around its mean $\sqrt{2N}$
and fluctuates over a scale of width $N^{-1/6}$. The probability of fluctuations
on this scale is described by the Tracy-Widom distribution (shown schematically).}
\label{figtw}
\end{figure}

Studies of the statistics of the eigenvalues of random matrices have a
long history going back to the seminal work of Wigner~\cite{Wigner}.
Since then, random matrices have found applications in multiple fields
including nuclear physics, quantum chaos, disordered systems, string
theory and number theory~\cite{Mehta}. Three classes of matrices with
Gaussian entries have played important roles~\cite{Mehta}: $(N\times
N)$ real symmetric (Gaussian Orthogonal Ensemble (GOE)), $(N\times N)$
complex Hermitian (Gaussian Unitary Ensemble (GUE)) and $(2N\times
2N)$ self-dual Hermitian matrices (Gaussian Symplectic Ensemble
(GSE)).  A central result in the theory of random matrices is the
celebrated Wigner semi-circle law. It states that for large $N$ and
on an average, the $N$ eigenvalues (suitably scaled) lie within a
finite interval $\left[-\sqrt{2N}, \sqrt{2N}\right]$, often referred
to as the Wigner `sea'. Within this sea, the average density of states
has a semi-circular form (see Fig. \ref{figtw}) that vanishes at the
two edges $-\sqrt{2N}$ and $\sqrt{2N}$
\begin{equation}
\rho_{\rm sc}(\lambda,N) = \sqrt{\frac{2}{N\pi^2}}\,{\left[1 -\frac{\lambda^2}{2N}\right]}^{1/2}.
\label{wig1}
\end{equation}

Thus, the average of the maximum (minimum) eigenvalue is $\sqrt{2N}$
(-$\sqrt{2N}$).  However, for finite but large $N$, the maximum
eigenvalue fluctuates, around its mean $\sqrt{2N}$, from one sample to
another. Relatively recently Tracy and Widom~\cite{TW} proved that
these fluctuations {\em typically} occur over a narrow scale of $\sim
O(N^{-1/6})$ around the upper edge $\sqrt{2N}$ of the Wigner sea (see
Fig. 1). More precisely, they showed~\cite{TW} that asymptotically for
large $N$, the scaling variable $\xi=\sqrt{2}\,N^{1/6}\, [\lambda_{\rm
max}-\sqrt{2N}]$ has a limiting $N$-independent probability
distribution, ${\rm Prob}[\xi\le x]= F_{\beta}(x)$ whose form depends
on the value of the parameter $\beta=1$, $2$ and $4$ characterizing
respectively the GOE, GUE and GSE. The function $F_{\beta}(x)$,
computed as a solution of a nonlinear differential equation~\cite{TW},
approaches to $1$ as $x\to \infty$ and decays rapidly to zero as $x\to
-\infty$. For example, for $\beta=2$, $F_2(x)$ has the following
tails~\cite{TW},
\begin{eqnarray}
F_2(x) &\to & 1- O\left(\exp[-4x^{3/2}/3]\right)\quad\, {\rm as}\,\,\, x\to \infty 
\nonumber \\  
&\to & \exp[-|x|^3/12] \quad\, {\rm as}\,\,\, x\to -\infty.
\label{asymp1}
\end{eqnarray}
The probability density function $dF_{\beta}/dx$ thus has highly
asymmetric tails.  The distribution of the minimum eigenvalue simply
follows from the fact that ${\rm Prob}[\lambda_{\rm min}\ge
\zeta]={\rm Prob}[\lambda_{\rm max}\le -\zeta]$.  Amazingly, the
Tracy-Widom distribution has since emerged in a number of seemingly
unrelated problems such as the longest increasing subsequence
problem~\cite{BDJ}, directed polymers in $(1+1)$-dimensions~\cite{DP},
various $(1+1)$-dimensional growth models~\cite{Growth}, 
a class of sequence alignment problems~\cite{MN} and in finance~\cite{BBP}. Recently, it has been
shown that the statistics of the largest eigenvalue is also of importance 
in population growth of organisms in fluctuating environments~\cite{KL}.

The Tracy-Widom distribution describes the probability of {\em typical and small}
fluctuations of $\lambda_{\rm max}$ over a very narrow region of width
$\sim O(N^{-1/6})$ around the mean $\langle \lambda_{\rm max}\rangle
\approx \sqrt{2N}$. A natural question is how to describe the
probability of {\em atypical and large} fluctuations of $\lambda_{max}$ around its
mean, say over a wider region of width $\sim O(N^{1/2})$? For example,
what is the probability that all the eigenvalues of a random matrix
are negative (or equivalently all are positive)?  This is the same as
the probability that $\lambda_{\rm max}\le 0$ (or equivalently
$\lambda_{\rm min}\ge 0$). Since $\langle \lambda_{\rm max}\rangle
\approx \sqrt{2N} $, this requires the computation of the probability
of an extremely rare event characterizing a large deviation of $\sim
-O(N^{1/2})$ to the left of the mean. This question recently came up
in the context of random landscape models of antropic principle
based string theory~\cite{Susskind,AE} as well as in quantum
cosmology~\cite{MH}.  Here one is interested in the statistical
properties of vacua associated with a random multifield potential,
e.g., how many minima are there in a random string landscape? Similar
questions also arise in disordered systems where one is interested in
counting the number of local minima of a random Gaussian
field~\cite{Fyodorov}. In order to have a local minimum of the
random landscape one needs to ensure that the eigenvalues of the
associated random Hessian matrix are all positive. A related important
question is: if one conditions all the eigenvalues to be positive, how
does the average density of states get modified from the Wigner
semi-circle form? In this Letter, we address these issues
analytically.

It is useful to summarize our main results. In Ref.~\cite{AE}, it was
shown numerically that the probability that all the eigenvalues of a
$(N\times N)$ GOE matrix ($\beta=1$) are positive (or equivalently all
the eigenvalues are negative, i.e., $\lambda_{\rm max}\le 0$)
decreases rapidly with large $N$ as ${\rm Prob}[\lambda_{\rm max}\le
0] \sim \exp\left[-\theta(0) N^2\right]$.  A crude approximate
argument was provided for the exponent $\theta(0)\approx
1/4$~\cite{AE}, along with numerical simulations. Here we show exactly
that for all ensembles characterized by the parameter $\beta$,
\begin{equation}
\theta(0)= \beta \frac{\ln 3}{4} = (0.274653\dots )\,\beta.
\label{expo1}
\end{equation}
More generally we calculate the exact large deviation function associated with
large fluctuations of $\sim -O(N^{1/2})$ of $\lambda_{\rm max}$ to the left of its
mean value $\sqrt{2N}$. We show that for large $N$ and for all ensembles
\begin{equation}
{\rm Prob}\left[\lambda_{\rm max}\le t, N\right] \sim \exp\left[-\beta
N^2 \Phi\left( \frac{\sqrt{2N}-t}{\sqrt{N}} \right) \right]
\label{ldf1}
\end{equation}
where $t\sim O(N^{1/2})\le \sqrt{2N}$ is located deep inside the
Wigner sea. The large deviation function $\Phi(y)$ is zero for $y\le 0$,
but is nontrivial for $y> 0$ which we compute exactly.
For {\em small} deviations to the left of the mean, taking the 
$y\to 0$ limit of $\Phi(y)$, we recover the left tail of the Tracy-Widom distribution
as in Eq. (\ref{asymp1}). Thus our result for {\em large} deviations
of $\sim -O(N^{1/2})$ to the left of the mean is complementary to the
Tracy-Widom result for {\em small} fluctuations of $\sim -O(N^{-1/6})$
and the two solutions match smoothly.  In the process, we also
calculate exactly the modified average density of states when all the
eigenvalues are constrained to be on the right of a barrier say at
$\lambda=\zeta$, thus generalizing  Wigner's semi-circle law.

Our starting point is the celebrated result due to Wigner for the
joint probability density function (pdf) of the eigenvalues of 
a random $(N\times N)$ matrix~\cite{Mehta} 
\begin{equation}
P(\{\lambda_i\}) = B_N \exp\left[-\frac{\beta}{2}\left(\sum_{i=1}^N\lambda_i^2 
-\sum_{i\ne j}\ln(|\lambda_i-\lambda_j|)\right)\right],
\label{pdf}
\end{equation}
where $B_N$ normalizes the pdf and $\beta=1$, $2$ and $4$ correspond
respectively to the GOE, GUE and GSE.  The joint law allows one to
interpret the eigenvalues as the positions of charged particles,
repelling each other via a $2$-d Coulomb potential (logarithmic);
they are confined on a $1$-d line and each is subject to an external harmonic
potential. The parameter $\beta$ that characterizes the type of 
ensemble can be interpreted as the inverse temperature. The average
density of states $\rho_{\rm sc}(\lambda,N)= \sum_{i=1}^N\langle
\delta(\lambda-\lambda_i)\rangle/N$ can be calculated~\cite{Mehta}
from the joint pdf in Eq. (\ref{pdf}) and has the Wigner
semi-circular form of Eq. (\ref{wig1}). In the Coulomb gas language,
this is the average equilibrium charge density.

Here we are interested in the probability $Q_N(\zeta)$ that all the
eigenvalues are bigger than say $\zeta$, i.e., the probability that
all charges lie to the right of the barrier at $\zeta$. Note that, due
to the $\lambda\to -\lambda$ symmetry of the pdf in Eq. (\ref{pdf}),
this is also the probability that all eigenvalues are less than
$-\zeta$, i.e., the probability that $\lambda_{\rm max}\le
-\zeta$. Let us first define the restricted partition function
\begin{eqnarray}
Z_N(\zeta)&=& \nonumber \int_{\lambda_i >\zeta}^\infty\prod_{i=1}^N
d\lambda_i \nonumber \\ & &\exp\left[-\frac{\beta}{2}
\left(\sum_{i=1}^N \lambda_i^2 -\sum_{i\ne j}\ln
\left(|\lambda_i-\lambda_j|\right) \right) \right].
\label{rpf}
\end{eqnarray}
It then follows that
\begin{equation}
Q_N(\zeta) = \frac{Z_N(\zeta)}{Z_N(-\infty)}.
\label{eqtd}
\end{equation}

Let  $\rho_N(\lambda)=\sum_{i=1}^N
\delta(\lambda-\lambda_i)/N$ denote the spatial density of charges.
Using standard techniques of functional integration we may express
$Z_N(\zeta)$ as~\cite{prep}
\begin{eqnarray}
&&Z_N(\zeta) \propto \int {\cal D}[\rho_N] \exp \left[ -\frac{\beta
    N}{2} \int_\zeta^\infty d\lambda \ \rho_N(\lambda) \lambda^2
  \right. \nonumber \\ &+&\frac{\beta N^2}{2} \int_{\zeta}^{\infty}
  d\lambda\, d\lambda' \ \rho_N(\lambda)\rho_N(\lambda')
  \ln\left(|\lambda -\lambda'|\right)\nonumber \\ &-&\left. { N}
  \int_{\zeta}^{\infty} d\lambda \ \rho_N(\lambda)\ln\left(
  \rho_N(\lambda)\right) \right].
\label{fi1}
\end{eqnarray}
where the first two terms represent the energy of the charges as in
Eq. (\ref{rpf}).  The third term represents the entropy which has a
mean field form due to the fact that all charges interact with each
other via the long-range logarithmic potential.  The charge density
$\rho_N(\lambda)$ evidently satisfies the constraints:
$\rho_N(\lambda) = 0\ {\rm for}\ \lambda < \zeta$ and
$\int_\zeta^\infty d\lambda \rho_N(\lambda) = 1$.

Since we are interested in fluctuations of $\sim O(N^{1/2})$, 
it is convenient to work with the rescaled variables, 
$\lambda = \mu\sqrt{N}$ and $\zeta = z\sqrt{N}$. 
It is reasonable to assume that the charge density scales as,
$\rho_N(\lambda)= N^{-1/2} f\left(\lambda N^{-1/2}\right)$.
The scaling function evidently satisfies the constraints:  
\begin{equation}
\int_z^\infty d\mu  f(\mu) = 1 ;\ \ \ f(\mu) = 0\ {\rm for}\  \mu < z.
\label{eqnf}
\end{equation}
Expressing the action in Eq. (\ref{fi1}) in terms of rescaled charged
density $f(\mu)$, one finds that the energy term scales as $\sim
O(N^2)$ whereas the entropy term $\sim O(N)$ is subdominant for large
$N$. For large $N$, the functional integration can be carried out
using the method of steepest descent. This gives, as a function of
rescaled variable $z=\zeta/\sqrt{N}$,
\begin{equation}
Z_N(z) \propto \exp\left[\beta N^2 S(z) + O(N)\right]
\label{eqst}
\end{equation}
where $ S(z) = \max_f\left\{ \Sigma(f)\right\}$ 
and 
\begin{eqnarray}
\Sigma(f) &=& 
-\frac{1}{2}\int_z^\infty d\mu \ f(\mu) \mu^2 \nonumber \\ &+&
\frac{1}{2}
\int_z^\infty \int_z^{\infty}d\mu d\mu' f(\mu) f(\mu') \ln\left(|\mu-\mu'|\right).
\label{sigmaf} 
\end{eqnarray}

\begin{figure}
\includegraphics[width=.9\hsize]{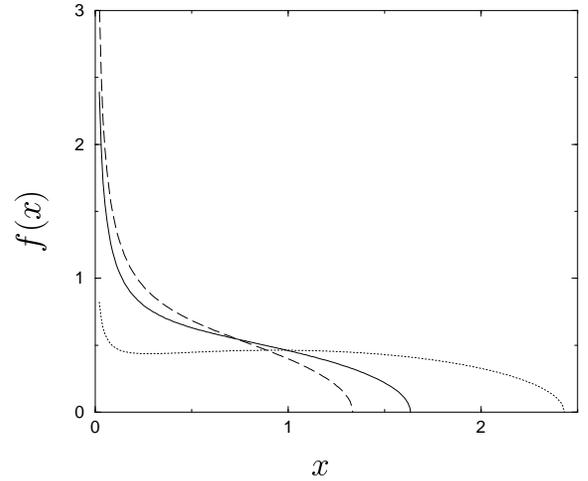}
\caption{The average density of states $f(x)$ plotted as a function of the 
shifted variable $x$ for $z=-1$ (dotted line), $z= 0$ (solid line), and $z=0.5$ 
(dashed line). }
\label{figfx}
\end{figure}

The stationarity condition $\delta \Sigma(f)/{\delta f}=0$ gives
\begin{equation}
\frac{\mu^2}{2} + C = 
\int_z^\infty d\mu'\ f(\mu') \ln\left(|\mu-\mu'|\right),
\label{eqvar}
\end{equation}
where $C$ is a Lagrange multiplier enforcing the normalization of $f$
in Eq. (\ref{eqnf}). Differentiating 
Eq. (\ref{eqvar}) with respect to $\mu$ gives
\begin{equation}
\mu = {\cal P}\int_z^\infty d\mu'\ f(\mu') \frac{1}{\mu -\mu'},
\label{eqdvar}
\end{equation}
where $\cal{P}$ indicates the Cauchy principle part. It is convenient to
introduce a shift $\mu=z+x$ 
where $x\ge 0$ represents the distance from the barrier (to the right) at $z$.
In terms of the variable $x$, Eq. (\ref{eqdvar}) becomes an integral equation
for the charge density 
\begin{equation}
x + z = {\cal P}\int_0^\infty dx'\ f(x') \frac{1}{x -x'}
\label{sht}
\end{equation}
where the rhs represents a semi-infinite Hilbert transform. The real
technical challenge is to invert this integral equation and obtain
a closed form expression for the rescaled charge density $f(x)$. Fortunately
this can be done~\cite{prep}. We
find that $f(x)$ is nonzero inside a finite box $x\in [0, L(z)]$ and vanishes
outside this box. For $0\le x\le L(z)$, the density is given exactly by
\begin{equation}
f(x)= \frac{1}{2\pi \sqrt{x}}\,\sqrt{L(z)-x}\, \left[L(z)+2x + 2z\right].
\label{eqfz}
\end{equation}
The length of the box $L(z)$ can be determined from the normalization condition
in Eq. (\ref{eqnf}) and is given by
\begin{equation}
L(z) = \frac{2}{3} \left[ \sqrt{z^2 + 6} -z\right].
\label{lz}
\end{equation}
Note that the charge density $f(x)$ depends
on $z$, i.e. the location of the barrier. A plot of this density  
for several values of $z$ are shown in Fig. \ref{figfx}.

\begin{figure}
\includegraphics[width=.9\hsize]{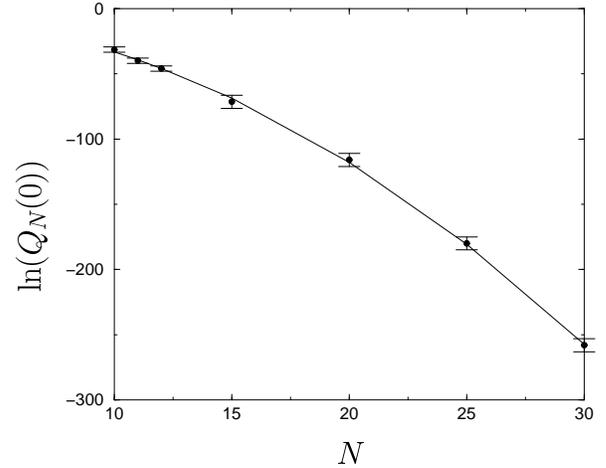}
\caption{Monte Carlo computation of $\ln(Q_N(0))$ points with error bars 
along with a quadratic fit (solid line).}
\label{figtheta}
\end{figure}

A couple of remarks are in order: (i) the charge density $f(x)$ must
be positive for all $x$ including $x=0$. As $x\to 0$, $f(x)$ diverges
as $x^{-1/2}$. However in order that it remains positive, we need to
ensure that the amplitude $L(z)+2z\ge 0$ at $x=0$ in
Eq. (\ref{eqfz}). This condition, using $L(z)$ from Eq. (\ref{lz}),
requires $z\ge -\sqrt{2}$. Thus the results in Eqs. (\ref{eqfz}) and
(\ref{lz}) are valid only for $z\ge -\sqrt{2}$. Indeed, this is
expected because exactly at $z=-\sqrt{2}$, i.e., when the barrier is
placed at the left edge of the Wigner sea, we recover from
Eq. (\ref{eqfz}) the Wigner semi-circle law. For $z=-\sqrt{2}$,
Eq. (\ref{lz}) gives $L=2\sqrt{2}$ (the support of the semi-circle)
and Eq. (\ref{eqfz}) gives $f(\mu)= \sqrt{2-\mu^2}/\pi$ for
$-\sqrt{2}\le \mu\le \sqrt{2}$. Thus, for any $z<-\sqrt{2}$, our exact
solution indicates that the charge density remains unchanged from the
Wigner semi-circular form. Physically this means that if the wall is
placed to the left of the lower edge of the Wigner sea, it has no
effect on the charge distribution. (ii) The second remark is that the
charge density $f(x)$ changes its shape in an interesting fashion as
one changes the barrier location $z$ (see Fig. {\ref{figfx}).  It
turns out that for any $z> -\sqrt{2}$, the charges always accumulate
near the barrier at $x=0$ leading to a square-root divergence of
$f(x)\sim x^{-1/2}$ as $x\to 0$. In particular, for $z=0$, this
accumulation of eigenvalues near $x=0$ can be interpreted as the
accumulation of massless modes in the context of a (stable) field
theory, a fact that may be of relevance in anthropic principle based
string theory.

\begin{figure}
\includegraphics[width=.9\hsize]{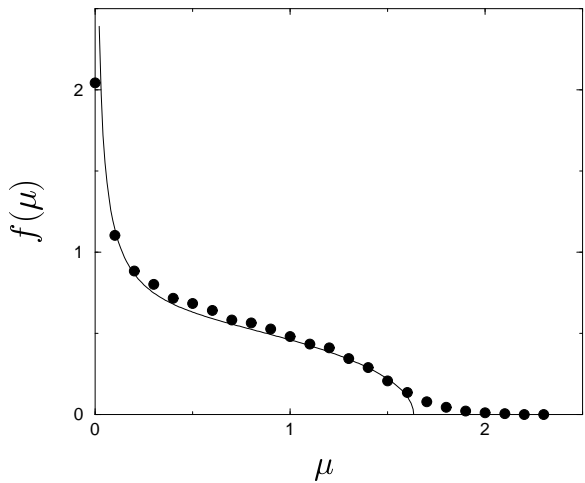}
\caption{The analytic large $N$ formula for $f$ with $z=0$ (solid
line) in Eq. (\ref{eqfz}) is compared to the numerically generated
averaged histogram of $(6\times 6)$ Gaussian matrices with positive
eigenvalues. Despite the small size $N=6$, the agreement is
already fairly good, except near the large $\mu$ tail.}
\label{figrhoz}
\end{figure}

Knowing $f(x)$ exactly, the Lagrange multiplier is determined
by setting $\mu=z$ in Eq. (\ref{eqvar}). This gives, following
a shift in the integral,
$C=-{z^2/2}+ \int_0^\infty dx'\ f(x') \ln\left(x'\right)$.
The saddle point action can now be evaluated explicitly~\cite{prep}
\begin{eqnarray}
& &S(z) = -\frac{1}{216}\left[ 72z^2 -2z^4 +(30z + 2z^3) \sqrt{6+z^2}
\right.\nonumber \\ &+&\left. 27\left( 3 + \ln(1296) - 4 \ln\left(-z +
\sqrt{6 +z^2}\right) \right) \right].
\label{action}
\end{eqnarray} 
The probability that all eigenvalues are to the right of $\zeta =z \sqrt{N}$
is then given by, to leading order in large $N$, using Eqs. (\ref{eqst}) and ({\ref{eqtd})
\begin{equation}
Q_N\left(\zeta=z\sqrt{N}\right) \approx \exp\left[-\beta N^2
\theta(z)\right]
\label{asymp2}
\end{equation}
where $\theta(z)= S(-\sqrt{2})-S(z)$ and $S(z)$ is given by
Eq. (\ref{action}). Note that we have used $S(-\infty)=S(-\sqrt{2})$
following remark (i) above.  The result in Eq. (\ref{ldf1}) can then
be derived by setting $t=-\zeta=-z\sqrt{N}$ and one finds the large
deviation function for $y\ge 0$, $\Phi(y)= S(-\sqrt{2})-S(-\sqrt{2}+y)$. For small
$y$, $\Phi(y) \approx y^3/6\sqrt{2}$ and for large $y$,
$\Phi(y)\approx y^2/2$.  Thus for $\sqrt{2N}-t <<
\sqrt{N}$, using $\Phi(y)\approx y^3/{6\sqrt{2}}$ we get,
\begin{equation}
{\rm Prob}[\lambda_{\rm max}\le t, N]\approx \exp\left[-\frac{\beta}{24}\big|\sqrt{2}\, 
N^{1/6}\,(t-\sqrt{2N})\big|^3\right]
\label{asymp3}
\end{equation}
which matches exactly with the left tail of the Tracy-Widom distribution for all $\beta$.
For example, for $\beta=2$ one can easily verify this by comparing Eqs. (\ref{asymp3})
and (\ref{asymp1}).

The probability that all eigenvalues are positive is obtained by
setting $z=0$ in Eq. (\ref{asymp2}) resulting in a remarkably simple
and exact formula stated in Eq. (\ref{expo1}). The fact that this
probability decreases as rapidly as $\sim \exp[-\beta \theta(0) N^2]$
for large $N$ and that there are significant $\sim O(N)$ corrections
indicate that numerically it is extremely difficult to measure the
exponent $\theta(0)$ accurately.  An attempt was made in
Ref.~\cite{AE} using  GOE ($\beta=1$) matrices up to sizes of 
$N=7$ to fit the probability with the form $\exp[-a N^{\alpha}]$ 
that yielded $\alpha \approx 2.00387$ and $a\approx 0.3291$.  
Clearly, the system sizes are
too small to take this fit seriously.  It turns out that instead it is
easier to evaluate $Q_N(0)$ directly from Eq. (\ref{eqtd}) via a
clever Monte Carlo method that allows us to go up to $N\sim
30$~\cite{prep}.  In Fig. \ref{figtheta} we show a plot of
$\ln(Q_N(0))$ measured using this Monte Carlo method (for $\beta=1$)
with a fit of the form $aN^2 + b N +c$. This fit yields $a \approx
-0.2755$ which is in good agreement with the exact value of
$\theta(0)=0.274653..$ predicted here.

Another numerical check consists in computing the charge density
$f(\mu)$ by direct sampling of Gaussian matrices and comparing it to
the theoretical prediction in Eq. (\ref{eqfz}). Here, we are clearly
restricted to small values of $N$.  In Fig. \ref{figrhoz}, we compare
the numerically computed $f(\mu)$ for $z=0$ obtained from matrices of
size $(6\times 6)$ with the theoretical prediction.  Despite the small
value of  $N$, the agreement is already fairly good.

We thank O. Bohigas for useful comments.

\end{document}